\documentclass[usenatbib]{mn2e}
\usepackage{graphicx,tabulary,amsmath,amssymb,upgreek,wasysym,overpic,array,hyperref}
\usepackage[T1]{fontenc}

\newcommand{\changes}[1]{{#1}}

\numberwithin{equation}{section}
\title[Synthetic observations of protostellar multiples]{Synthetic observations of protostellar multiple systems}
\author[O. Lomax \& A. P. Whitworth]{O. Lomax\thanks{E-mail: oliver.lomax@astro.cf.ac.uk}, \& A. P. Whitworth\\
School of Physics and Astronomy, Cardiff University, Cardiff CF24 3AA, UK}

\begin{document}
\pagerange{\pageref{firstpage}--\pageref{lastpage}} \pubyear{2015}
\maketitle
\label{firstpage}

\begin{abstract}

\noindent Observations of protostars are often compared with synthetic observations of models in order to infer the underlying physical properties of the protostars. The majority of these models have a single protostar, attended by a disc and an envelope. However, observational and numerical evidence suggests that a large fraction of protostars form as multiple systems. This means that fitting models of single protostars to observations may be inappropriate.
We produce synthetic observations of protostellar multiple systems undergoing realistic, non-continuous accretion. These systems consist of multiple protostars with episodic luminosities, embedded self-consistently in discs and envelopes. We model the gas dynamics of these systems using smoothed particle hydrodynamics and we generate synthetic observations by post-processing the snapshots using the \textsc{spamcart} Monte Carlo radiative transfer code. 
We present simulation results of three model protostellar multiple systems. For each of these, we generate $4\times10^4$ synthetic spectra at different points in time and from different viewing angles. We propose a Bayesian method, using similar calculations to those presented here, but in greater numbers, to infer the physical properties of protostellar multiple systems from observations. 
\end{abstract}


\section{Introduction}


Stars form when dense prestellar cores in molecular clouds collapse under their own self-gravity. These cores are turbulent \citep[e.g.,][]{ABMP07} and therefore have net angular momentum. From conservation of angular momentum, any protostar which forms during the initial core collapse is likely to be attended by a circumstellar disc. As the remainder of the core envelope collapses inwards, additional material accretes onto the disc. Dissipative phenomena such as viscosity \citep[e.g.,][]{SW08} and magnetic braking \citep[e.g.,][]{Z07} transport angular momentum to the outer the regions of the disc, allowing material in the inner regions to accrete onto the protostar.

Further protostars may from via disc fragmentation if the following two criteria are met. First, the surface density of the disc must be great enough for self-gravity to overcome thermal and centrifugal support \citep{T64}. Second, the cooling time of a potential fragment must be shorter than its orbital period \citep{Gam01}. The timescale on which disc fragmentation occurs is very short \citep[$t_\textsc{frag}\sim10^4\,\mathrm{years}$; e.g.,][]{SW08} which makes observing the process difficult. However, observations of young protostellar multiples show separations consistent with disc fragmentation \citep[e.g.,][]{TLL16}. Furthermore, numerical simulations of disc fragmentation are able to reproduce the mass distribution and multiplicity statistics of low mass stars and brown dwarfs \citep[e.g.,][]{SW09a,LWHSW14,LWHSW14b,LWH15}. 

Accretion luminosity is the dominant source of radiative feedback for young protostars. Calculations which include continuous radiative feedback from accretion onto protostars produce discs which are too hot to fragment \citep[e.g.,][]{B09b,K06,KCKM10,OKMK09,O10}. This supports the hypothesis that disc fragmentation does not occur often. However, this hypothesis is weakened by the observed luminosities of young protostars. These are much lower than those predicted by continuous accretion models. This is known as the \emph{luminosity problem} \citep[first noted by][]{KHSS90}. This problem is alleviated by observational evidence suggesting that protostellar luminosities are episodic. For example, FU Ori stars exhibit luminosity outbursts which last of order decades \citep[e.g.,][]{HK96,GAR08,PSMB10}. Similarly, knots have been observed in protostellar outflows which have spacings suggestive of episodic accretion \citep[e.g.,][]{R89}.  Simulations which include sub-grid models of episodic accretion demonstrate that significant disc fragmentation may still occur if radiative feedback is episodic \citep[e.g.,][]{SWH11,SWH12}.


Inferring the physical properties of young protostars requires matching their observables, e.g. their spectral energy distributions (SEDs), with those of models. Great efforts have been made to construct large catalogs of SEDs from model protostars with realistic dust properties and varied protostar and disc parameters \citep[e.g.,][]{RWI06}. Additionally, previous work has modelled the observable features of embedded stars with variable luminosity \citep[e.g.,][]{H11,JHH13}. However, in all these cases the stars are single, and there is growing observational and theoretical evidence that a large fraction of stars form as multiples \citep[e.g.,][]{KIM11,HWGW13,B14,LWHSW14b}. Here, we simulate the observational signatures of embedded protostellar multiple systems via a two-stage process. First, we use smoothed particle hydrodynamics (SPH) to simulate the collapse and fragmentation of  turbulent prestellar cores. This uses an approximate on-the-fly radiative transfer method and sink particles which have episodic luminosities determined by a sub-grid model involving the magneto-rotational instability \citep{BH91}. Second, we post-process snapshots from the simulations using a Monte Carlo radiative transfer algorithm to calculate the dust emission and scattered light.

In Section \ref{sec:numerical_method} we detail the the numerical methods used to perform the calculations. In Section \ref{sec:sph_simulations} we describe the initial conditions and results of the SPH simulations. In Section \ref{sec:radiative_transfer} we define the parameters of the radiative transfer calculations and describe their results. In Section \ref{sec:discussion} we discuss the significance of these calculations, and propose a methodology for using them to interpret the properties of protostars from observations. Finally, we summarise our findings in Section \ref{sec:summary}.

\section{Numerical method}
\label{sec:numerical_method}

\changes{In this section, we outline the numerics of this work. Some of the more technical issues are included as appendices. If the reader is unconcerned with these details, we suggest they skip to Section \ref{sec:sph_simulations}.}

\subsection{Smoothed Particle Hydrodynamics}
\label{sec:sph}

\begin{figure*}
	\centering
	\includegraphics[width=\textwidth]{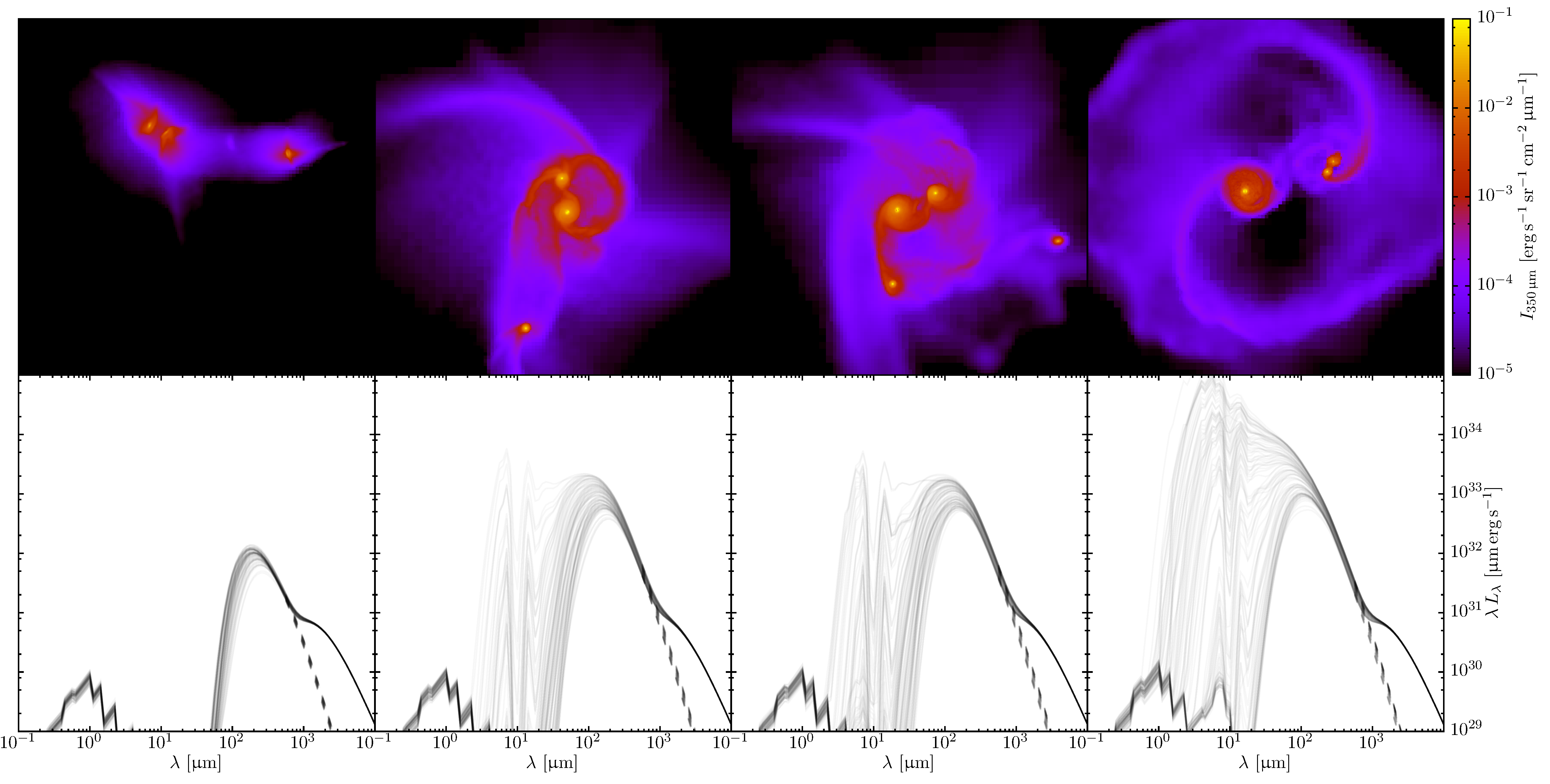}
	\caption{Synthetic observations of Core A. From left to right, the columns show snapshots at $t=1.0,\ 1.5,\ 1.7\ \text{and}\ 3.0\times10^{4}\,\mathrm{yrs}$. The top row shows the unconvolved $350\,\mathrm{\upmu m}$ intensity integrated along a fixed line of sight. The edge-length of each frame is $1300\,\mathrm{au}$. The bottom row shows an ensemble of source SEDs, seen through one hundred randomly chosen lines of sight. The solid lines show the full SEDs and the dashed segments show the SEDs after the CMB has been subtracted. \changes{The short wavelength component of the SEDs ($0.3\,\mathrm{\upmu m}\lesssim\lambda\lesssim3\,\mathrm{\upmu m}$) is predominantly scattered light from the interstellar radiation field.}}
	\label{fig:core_a_obs}
\end{figure*}

Core evolution is simulated using the \textsc{seren} $\nabla h$-SPH code \citep{HBMW11}, with $\eta = 1.2$ and the \citet{MM97} formulation of time dependent artificial viscosity. SPH particles have mass $m_\textsc{sph}=10^{-5}\,\mathrm{M}_{\odot}$ and the minimum mass for star formation \citep[$\sim3\times10^{-3}\,\mathrm{M}_{\odot}$; see][]{WS06} is resolved with $\sim300$ particles. Gravitationally bound regions with densities higher than $\rho_\textsc{sink}=10^{-9}\,\mathrm{g}\,\mathrm{cm}^{-3}$ are replaced with sink particles. These use the \textsc{NewSinks} smooth accretion algorithm \citep{HWW13}. Sink particles have radius $r_\textsc{sink}\simeq0.2\,\mathrm{au}$, corresponding to the smoothing length of an SPH particle with density equal to $\rho_\textsc{sink}$. The equation of state and the energy equation are treated with the radiative transfer approximation described by \citet{SWBG07} (hereafter SW07).

Episodic radiative feedback from sinks is also included. Each sink has a variable luminosity which follows the sub-grid episodic accretion model described in \citet[][]{SWH11} (hereafter SWH11). \changes{Here, the mass of a sink is split into its stellar mass $M_\star$ and a mass attributed to an unresolved Inner Accretion Disc (IAD) $M_\textsc{iad}$. The total mass $M=M_\star+M_\textsc{iad}$ is used for all gravitational force calculations in the simulations. Mass which accretes onto the sink is added to $M_\textsc{iad}$. This mass is then transferred from the IAD to the star via a low quiescent accretion phase which is punctuated by intense episodic outbursts. These outbursts almost completely deplete the mass of the IAD. The mass transfer rate from the IAD to the star is used to calculate the luminosity of a sink particle. The parameters of this model (e.g., episode intervals and duration) are based on calculations involving the magneto-rotational instability by \citet{ZHG09,ZHG10} and \citet{Z10}. This produces sinks with low accretion luminosities which briefly increase by a couple of orders of magnitude every $\sim10^4$ years. The duration of each outburst is $\sim2\times10^2$ years.}

\subsection{Post-processing radiative transfer}

\subsubsection{Smoothed particle Monte Carlo radiative transfer}

We perform post-process dust radiative transfer calculations on simulation snapshots using the \textsc{spamcart} Monte Carlo code \citep[][hereafter LW16]{LW16}. The algorithm is a gridless adaptation of the \citet{Lucy99} method, designed to operate on smoothed particles instead of uniform density cells. Here, the gas particles from the simulation represent the dusty interstellar medium. \changes{The sink particles are treated as isotropic point sources with luminosities $L_\star$ given by the SWH11 model\footnote{\changes{We note that the assumption of isotropy is a simplification. In nature, the protostars have IADs which focus emergent radiation towards the protostellar poles. This anisotropy is further increased by the presence of protostellar jets, which are not modelled here. We comment on this further in Section \ref{sec:caveats}.}}. We assume each sink has a blackbody SED determined by $L_\star$ and an assumed radius $R_\star=3\,\mathrm{M_\odot}$}. In addition, we include the local interstellar radiation field calculated by \citet{PS08}. The sources emit luminosity packets, which propagate through the SPH density field until they escape the system. \changes{The dust properties of the particles are defined and discussed in Appendix \ref{sec:dust_properties}.}

The energy absorbed, per unit time, per unit mass of dust, for a particle is estimated by summing the optical depth contributions from luminosity packets which pass through the particle's smoothing kernel,
\begin{equation}
	\dot{A}_i\simeq\frac{\ell_j}{m_i}\sum\limits_j\kappa_{\lambda_j}\,\varsigma_{ij}\,,
	\label{eqn:dust_abs}
\end{equation}
Here $\ell_j$ is the luminosity attributed to packet $j$, $\lambda_j$ is its wavelength, $m_i$ is the dust mass of particle $i$, $\varsigma_{ij}$ is the column density along the trajectory of packet $j$ through particle $i$, and $\kappa_{\lambda_j}$ is the dust absorption opacity at $\lambda_j$.

The energy scattered, per unit time, per unit mass, per unit wavelength can be estimated by summing packets with $\lambda_j$ in the interval $[\lambda,\lambda+\mathrm{d}\lambda]$. Here
\begin{equation}
	\dot{S}_{i\lambda}\,\mathrm{d}\lambda\simeq\frac{\ell_j}{m_i}\sum\limits_{j,\mathrm{d}\lambda}\sigma_{\lambda_j}\,\varsigma_{ij}\,,
	\label{eqn:dust_sca}
\end{equation}
where $\sigma_{\lambda_j}$ is the dust scattering opacity at $\lambda_j$. Note that we assume dust scattering is isotropic. \changes{Additional features of the code, such as dust sublimation and dealing with very high opacities, are discussed in Appendicies \ref{sec:dust_sub} and \ref{sec:mrw}.}

We generate synthetic images and spectra of dust emission and scattered light by performing a ray trace for each pixel of a virtual camera. We identify the collection of particles which are intersected by a ray centred on and normal to the pixel. They are then sorted is descending order of distance from the pixel. The intensity (or surface brightness) $I_{\lambda n}$  at the pixel position is calculated iteratively from $i=1$ to $n$ where,
\begin{equation}
	I_{\lambda i}=I_{\lambda i-1}\,\exp(-\chi_{\lambda}\,\varsigma_i)+j_{\lambda i}\,[1-\exp(-\chi_{\lambda}\,\varsigma_i)] .
\end{equation}
Here, $I_{\lambda0}$ is the intensity of the background interstellar radiation field, $\chi_\lambda\equiv\kappa_\lambda+\sigma_\lambda$ is the dust mass extinction at $\lambda$ and $\varsigma_i$ is the column density along the ray through particle $i$. The source function $j_{\lambda i}$ is given by,
\begin{equation}
	j_{\lambda i}=\frac{1}{\chi_\lambda}\left[B_\lambda(T_i)\,\kappa_\lambda + \frac{\dot{S}_{\lambda i}}{4\uppi} \right]\,,
\end{equation}
where $B_\lambda(T)$ is the Planck function and $T_i$ is found by inverting the equation:
\begin{equation}
	\dot{A}_i=4\,\sigma_\textsc{sb}\,\bar{\kappa}_\textsc{p}(T_i)\,T_i^4\,.
\end{equation}
Here, $\sigma_\textsc{sb}$ is the Stefan-Boltzmann constant and $\bar{\kappa}_\textsc{p}(T)$ is the Planck mean mass absorption coefficient. Images are constructed using a quadtree of rays. We adaptively refine the image plane until \emph{every} particle has \emph{at least} one ray with an impact parameter shorter than its smoothing length. In most cases, the intensity maps refine down to length scales $\lesssim0.1\,\mathrm{au}$, which allows us to resolve the sublimation radii of most protostars. We note that these intensity maps can display visible pixelation in low intensity regions. This is a necessary trade-off; if the maximum resolution was applied over the entire area, most of the maps presented here would be made up of roughly $10^{9}$ pixels. Storing and processing this large a quantity of data is impractical, given that we want to generate a large number of multi-wavelength datacubes. Direct starlight from sink particles, attenuated by dust, is added to the image after calculating the dust emission and scattered light.

\subsection{Code benchmark}

\begin{figure*}
	\centering
	\includegraphics[width=\textwidth]{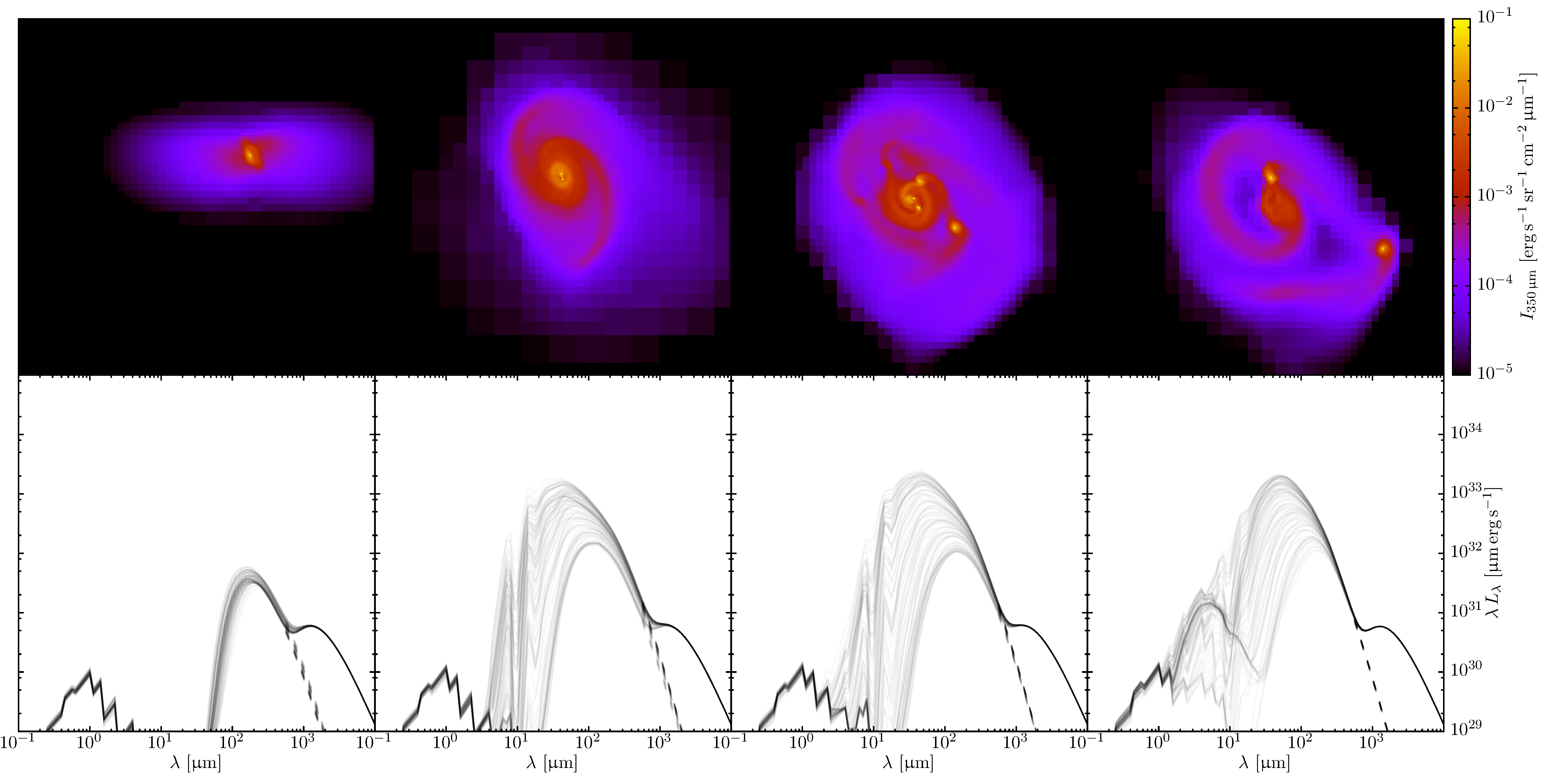}
	\caption{As Fig. \ref{fig:core_a_obs}, but for Core B at $t=2.0,\ 4.0,\ 5.0\ \text{and}\ 6.5\times10^{4}\,\mathrm{yrs}$. The edge length of each map is $700\,\mathrm{au}$.}
	\label{fig:core_b_obs}
\end{figure*}

\changes{Performing benchmarks for radiative transfer codes is often difficult. In most cases, multiple codes must be tested against each other in a labour-intensive study \citep[e.g.][]{BHW15,GBB17}. We acknowledge that this should be examined in future work. Here, instead, we perform a benchmark which demonstrates the code's ability to reproduce Kirchoff's law of thermal radiation. This benchmark is discussed in detail in Appendex \ref{sec:kirchoff}. We show that when an object (in this case, a non-spherically symmetric distribution of particles) reaches thermal equilibrium with a blackbody radiation field, the object is indistinguishable from the background. This applies for all viewing directions. When the background radiation field is that of a diluted blackbody, the object casts a silhouette against the background at short wavelengths, and glows relative to the background at long wavelengths.
}

\section{SPH simulations}
\label{sec:sph_simulations}

\subsection{Initial conditions}

We run three simulations of prestellar cores, one with the initial conditions taken from \citet{LWH15}, and two chosen from the ensemble generated by \citet{LWHSW14}. We apply a turbulent velocity field with a thermal mix of solenoidal to compressive modes and a $P(k)\propto k^{-4}$ power spectrum. The core density profile is that of a critical Bonner-Ebert sphere\footnote{\changes{We simply use the density profile of a critical Bonner-Ebert sphere. The cores are neither isothermal, nor in hydrostatic equilibrium.}}. The masses, sizes and non-thermal velocity dispersions of the cores are given in Table \ref{tab:core_params}\,. The core parameters are chosen because (i) they are similar to some of the observed cores in Ophiuchus, (ii) the masses roughly span the dynamic range observed in nearby young star forming regions and (iii) previous simulations, similar to these, produce more than one embedded protostar. We note that the setups are cherry-picked for their high multiplicities and the results should not be used to make statistical arguments. The simulations are terminated at $t=2\times10^5\,\mathrm{yrs}$, by which point accretion onto the protostars has ceased. This corresponds to the crossing time of cores in Ophiuchus, estimated by \citet{ABMP07}. 

\begin{table}
	\centering
	\begin{tabular}{cccc}
		\hline
		Core & $M_\textsc{core}\,\mathrm{[M_\odot]}$ & $R_\textsc{core}\,\mathrm{[au]}$ & $\sigma_\textsc{nt}\,\mathrm{[km\,s^{-1}]}$ \\
		\hline
		A & 2.9 & 3100 & 0.44 \\
		B & 1.0 & 3200 & 0.10 \\
		C & 0.5 & 2400 & 0.22 \\
		\hline
	\end{tabular}
	\caption{Core masses, $M_\textsc{core}$, radii, $R_\textsc{core}$ and non-thermal velocity dispersions, $\sigma_\textsc{nt}$. The parameters of Core A are taken from \citet{LWH15}. Cores B and C are chosen from the ensemble generated by \citet{LWHSW14}.}
	\label{tab:core_params}
\end{table}

\subsection{Results}
\label{sec:sph_results}
\begin{figure*}
	\centering
	\includegraphics[width=\textwidth]{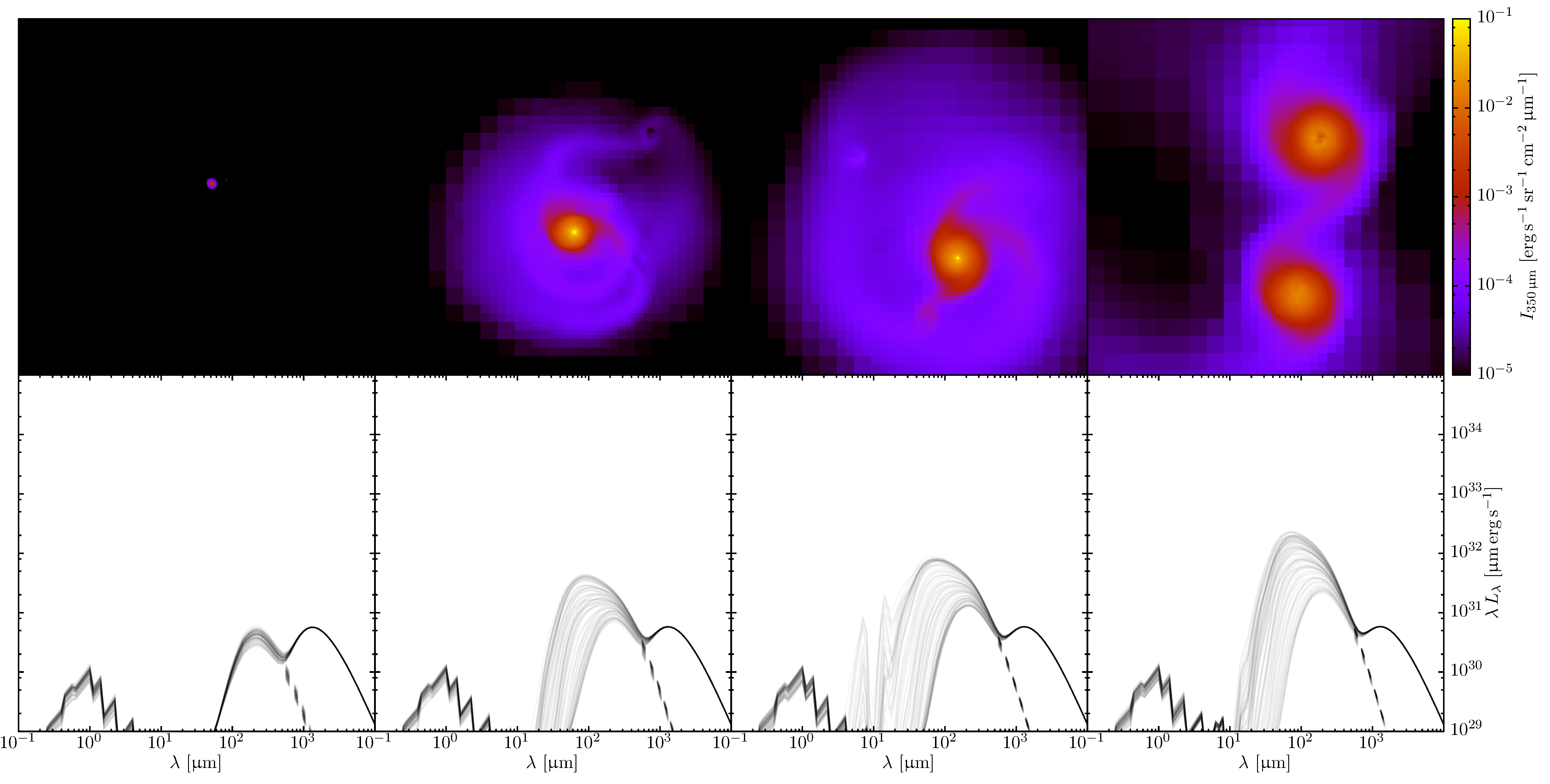}
	\caption{As Fig. \ref{fig:core_a_obs}, but for Core C at $t=1.7,\ 2.2,\ 2.3\ \text{and}\ 3.0\times10^{4}\,\mathrm{yrs}$. The edge length of each map is $300\,\mathrm{au}$.}
	\label{fig:core_c_obs}
\end{figure*}

Here we provide a brief overview of the evolution of each core. The number of protostars formed in each core simulation, along with their masses, is given in Table \ref{tab:protostars}\,. Intensity maps and SEDs highlighting the evolution Cores A, B and C are given in Figs. \ref{fig:core_a_obs}, \ref{fig:core_b_obs} and \ref{fig:core_c_obs} respectively.

\subsubsection{Core A}

Due to the strong turbulent velocity field, Core A fragments simultaneously into three protostars at $t\approx1.0\times10^4\,\mathrm{yrs}$. These quickly assemble into a triple system, attended by a circumsystem accretion disc. A fourth protostar forms via a gravitational instability in one of the accretion flows onto the disc at $t\approx1.5\times10^4\,\mathrm{yrs}$. The protostar quickly merges with the triple system to form a twin binary quadruple system. Additionally, gravitationally unstable material within the circumsystem disc fragments into a fifth object at $t\approx1.7\times10^4\,\mathrm{yrs}$. This object accretes very little mass and is ejected from the system at $t\approx3.0\times10^4\,\mathrm{yrs}$. The final system configuration is a twin binary quadruple system, composed of four proto-K-dwarfs, and an ejected proto-brown dwarf. By $t=2\times10^5\,\mathrm{yrs}$, 75\% of the original core mass has accreted onto the protostars.

\subsubsection{Core B}

Core B collapses into a single protostar at $t\approx3.0\times10^4\,\mathrm{yrs}$. A second protostar forms via disc fragmentation at $t\approx5.0\times10^4\,\mathrm{yrs}$, producing a binary system with a circumbinary disc. At $t\approx6.0\times10^4\,\mathrm{yrs}$, the disc fragments into a further five protostars. Two of these merge with the binary to form a twin binary quadruple of brown dwarfs. Later in the simulation, the other three objects are ejected as a single brown dwarf and a binary pair of brown dwarfs. By $t\approx2\times10^5\,\mathrm{yrs}$, 70\% of the original core mass has accreted onto the protostars.

\subsubsection{Core C}

Core C collapses into a single a protostar at $t\approx2.0\times10^4\,\mathrm{yrs}$. Three more protostars form via disc fragmentation between $t\approx2.2\ \text{and}\ t\approx2.3\times10^4\,\mathrm{yrs}$. The protostars settle into a twin binary quadruple (four brown dwarfs) by $t\approx3.0\times10^4\,\mathrm{yrs}$. By $t\approx2\times10^5\,\mathrm{yrs}$, 25\% of the original core mass has accreted onto the protostars.

\begin{table}
	\centering
	\begin{tabular}{ccl}
		\hline
		Core & $N_\star$ & $M_\star\,\mathrm{[M_\odot]}$ \\
		\hline
		A & 5 & 0.68, 0.56, 0.54, 0.44, 0.02 \\
		B & 7 & 0.44, 0.09, 0.07, 0.06, 0.05, 0.01, 0.01 \\
		C & 4 & 0.04, 0.03, 0.02, 0.02 \\
		\hline
	\end{tabular}
	\caption{The number of protostars, $N_\star$, and their masses, $M_\star$, formed in each core simulation.}
	\label{tab:protostars}
\end{table}

\section{Post-processing radiative transfer}
\label{sec:radiative_transfer}

\subsection{Calculation parameters}

\begin{figure*}
	\centering
	\includegraphics[width=\textwidth]{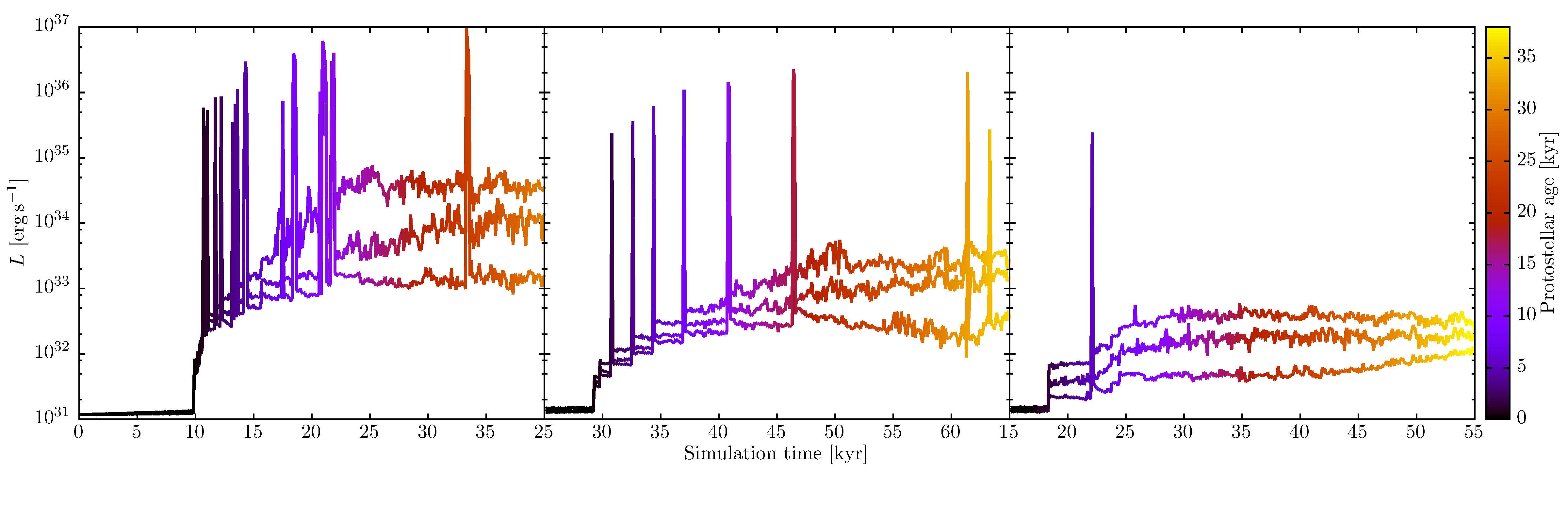}
	\caption{Apparent luminosities of Core A, Core B and Core C (left to right) as a function of simulation time. The three lines show the 15th, 50th and 85th centile values, seen through one hundred randomly chosen lines of sight. The colour scale shows the protostellar age, i.e. the age of the system where $t=0$ marks the formation of the first sink particle.}
	\label{fig:time_series}
\end{figure*}

For each core, we post-process 400 snapshots linearly spaced in time with an interval of 100 years. The ranges of these time series are shown by the $x$-axes in Fig. \ref{fig:time_series}. Each time series is limited to $4\times10^4\,\mathrm{yrs}$ in order to reduce computational expense.

For each of these calculations we inject $10^6$ equal-luminosity packets into the system from point sources, and $10^6$ luminosity packets into the system from the external radiation field. We iterate the calculation five times, after which the total luminosity from dust, i.e. $\sum_i A_i\,m_i$ for all particles, has converged to within a few percent.

For each post-processing calculation, we generate intensity maps from 100 random viewing directions, at 100 wavelengths equally spaced logarithmically in the range $10^{-1}\,\mathrm{\upmu m}\leq\lambda\leq10^{4}\,\mathrm{\upmu m}$. Each intensity map covers a $1500\,\mathrm{au}$ square centred on the particle ensemble's centre of mass. We generate SEDs for each line of sight by integrating over the image plane at each wavelength.

\subsection{Results}

\begin{figure*}
	\centering
	\includegraphics[width=\textwidth]{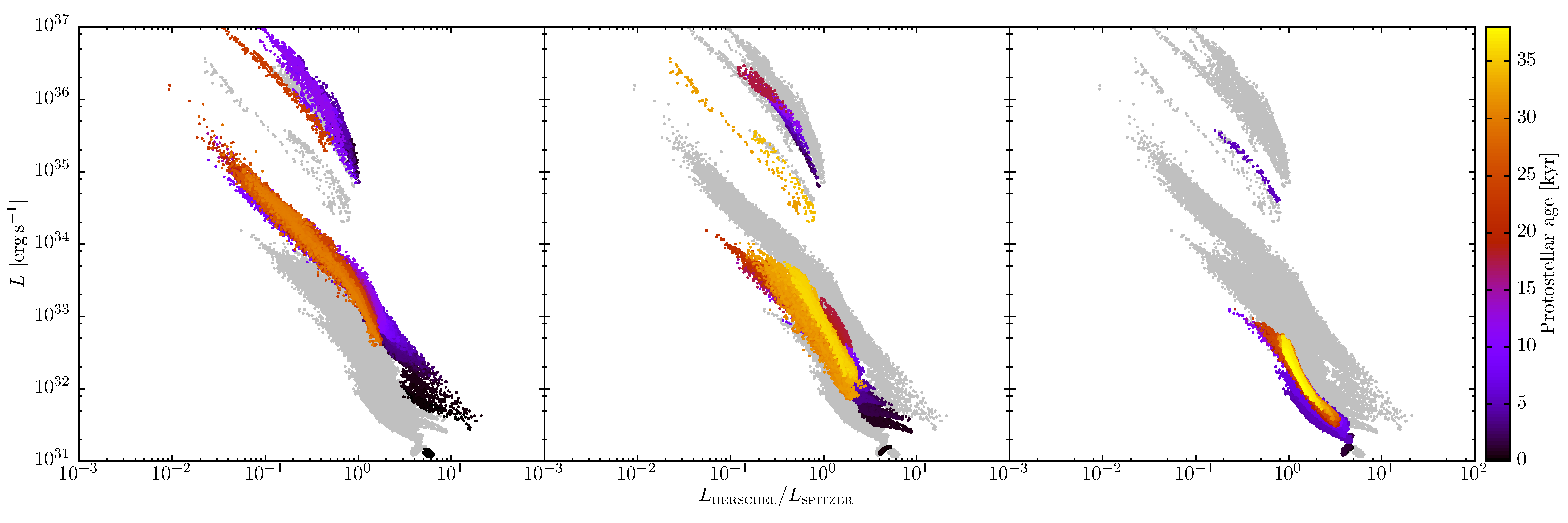}
	\caption{Apparent luminosities of Core A, Core B and Core C (left to right) plotted against infrared colour. Here, we define colour as the amount of luminosity in a Herschel-like band ($55\,\mathrm{\upmu m}\leq\lambda\leq672\,\mathrm{\upmu m}$) over that in a Spitzer-like band ($3\,\mathrm{\upmu m}\leq\lambda\leq180\,\mathrm{\upmu m}$). In each plot, the coloured points show the data related to the specific core. Data for the remaining two cores are shown in grey to provide a comparison. The colour scale shows the protostellar age and is the same as in Fig. \ref{fig:time_series}\,.}
	\label{fig:colour_magnitude}
\end{figure*}

\subsubsection{Photometry}

Fig. \ref{fig:time_series} shows the apparent luminosity of Core A, Core B and Core C as a function of time, where
\begin{equation}
	\begin{split}
		L&=\int\limits_0^\infty L_\lambda \mathrm{d}\lambda\,;\\
		L_\lambda&=4\,\uppi\int\limits_A I_\lambda\,\mathrm{d}A\,.
	\end{split}
\end{equation}
Here $A$ is the surface area of the map, which is equivalent to the solid angle subtended on the sky, multiplied by the distance to the source squared. Note that $L$ varies with viewing angle unless the source is isotropic or completely optically thin. The three lines in each plot of Fig. \ref{fig:time_series} show the 15th, 50th and 85th centile values of $L$ from multiple viewing angles as a function of time. \changes{The large peaks every $\sim10^4\,\mathrm{yrs}$ show the episodic outburst events. In general, these are much more frequent from more massive protostars. Between accretion bursts, the apparent luminosity depends strongly on viewing angle. The sources can easily appear 3 to 10 times more or less luminous than they really are. During an outburst, the variation with viewing angle is greatly reduced. This is because the increased stellar luminosity clears out a sublimation radius of $\sim2\,\mathrm{au}$ (see Appendix \ref{sec:dust_sub}) around the star, allowing radiation to escape easily in many directions. We note that in most cases, disentangling the apparent luminosity of each individual star is difficult as the direct stellar radiation is almost entirely reprocessed by the surrounding dust.}

The protostars in Core A settle to a median total luminosity\footnote{\changes{Here, the median value is similar to the direction-averaged mean value (i.e. the \emph{true} luminosity) to within a factor of a few. However, unlike the mean, the median is less sensitive to statistical outliers. In most cases, the median is less than the mean value.}} of $L\sim10^{34}\,\mathrm{erg\,s^{-1}}$, roughly $3\times10^4\,\mathrm{yr}$ after the formation of the first protostar ($L_\odot=3.8\times10^{33}\,\mathrm{erg\,s^{-1}}$). By this point, the four most massive protostars contribute approximately 40\%, 20\%, 20\% and 20\% of the total system luminosity. Over this time period, twelve episodic outbursts, each lasting $\lesssim10^2\,\mathrm{yrs}$, temporarily raise the system luminosity by three orders of magnitude. The protostars in Core B settle to a median total luminosity of $L\sim10^{33}\,\mathrm{erg\,s^{-1}}$ after $3\times10^4\,\mathrm{yr}$. Here, the six most massive stars contribute 70\%, 13\%, 8\%, 4\%, 3\% and 2\% of the total luminosity. Eight episodic outbursts occur during this period. The protostars in Core C reach a median total luminosity of $L\sim10^{32}\,\mathrm{erg\,s^{-1}}$ within $10^4\,\mathrm{yr}$. Here, the four most massive protostars contribute approximately 40\%, 20\%, 20\% and 20\% of the total luminosity. Only one episodic outburst occurs during this period.


Fig. \ref{fig:colour_magnitude} shows the colour-luminosity diagram for each protostellar system, over all times and viewing angles. Here, we define colour as the amount of luminosity in a Herschel-like band ($55\,\mathrm{\upmu m}\leq\lambda\leq672\,\mathrm{\upmu m}$) over that in a Spitzer-like band ($3\,\mathrm{\upmu m}\leq\lambda\leq180\,\mathrm{\upmu m}$). We refer to objects with low values of $L_\textsc{herschel}/L_\textsc{spitzer}$ as blue, and those with high values as red.

During the prestellar phase, each core has a low luminosity and is relatively red. As protostars form within the cores, the system move onto a \emph{main track} where they appear more blue and luminous the closer they are to being viewed face on. \emph{Secondary tracks} appear above the main track during episodic outbursts. While all three cores share these features, Core A has a bluer and more luminous main track than Core B, which has a bluer and more luminous main track than Core C.


\subsubsection{Spectral features}

Figs. \ref{fig:core_a_obs}, \ref{fig:core_b_obs} and \ref{fig:core_c_obs} show $350\,\mathrm{\upmu m}$ specific intensity maps and SEDs for Cores A, B and C respectively. All of the SEDs display some common features. At wavelengths $10^3\,\mathrm{\upmu m}\leq\lambda\leq10^4\,\mathrm{\upmu m}$, the radiation is primarily from the cosmic microwave backgrounds (CMB). The radiation at wavelengths $10^{-1}\,\mathrm{\upmu m}\leq\lambda\leq10^1\,\mathrm{\upmu m}$ is mostly scattered light from the interstellar radiation field\footnote{Note that the jagged features in the SED are caused by the wavelength binning of scattered luminosity packets during the calculation. This bin size can be reduced at the cost of introducing greater statistical noise.}. In between these bands, the radiation is predominantly dust emission from the core envelope and/or protostellar discs. This varies strongly from source to source as well as over time and through different viewing angles. We note that in the majority of cases, luminosity packets from protostars undergo $10^4$ to $10^6$ absorption/reemission or scattering events before they escape the system. Therefore the SED shape of the protostellar photosphere makes almost no contribution to the full system SED.


During the prestellar and very early protostellar stages of the simulations, i.e., where a protostar has formed but accounts for $\lesssim10\%$ of the total mass, the dust emission from all three cores resembles a $[T\sim15\,\mathrm{K},\beta\sim2]$ modified blackbody. Here, the source is only bright at wavelengths $10^2\,\mathrm{\upmu m}\lesssim\lambda\lesssim10^3\,\mathrm{\upmu m}$. \changes{The SED is only weakly dependent on viewing angle because the protostars are still embedded in dense clumps with some degree of spherical symmetry. Short wavelength emission is degraded to longer wavelengths in all directions.} As the protostars evolve over the next $\sim10^4\mathrm{yrs}$ (during which secondary protostars also form via disc fragmentation), they heat nearby dust and become bright at wavelengths $10\,\mathrm{\upmu m}\lesssim\lambda\lesssim10^2\,\mathrm{\upmu m}$. As these systems are roughly coplanar, most of the radiation can only escape at angles close to the rotational axis of the system. Systems with sufficiently luminous embedded protostars also emit significant amounts of radiation at wavelengths $1\,\mathrm{\upmu m}\lesssim\lambda\lesssim10\,\mathrm{\upmu m}$. This is seen strongly in Core A and moderately in Core B.

The systems described here are naturally dynamic and their morphologies can change dramatically over the course of $\lesssim10^4\,\mathrm{yrs}$. For example, Core C quickly transitions from an unstable disc around a single protostar to a stable quadruple system. While these events do not necessarily produce an obvious tracer in the SEDs, they can occasionally produce some unusual spectra. In Core A (see the last frame of Fig. \ref{fig:core_a_obs}), the final quadruple system forms during an event where a relatively discless protostar destroys the disc of another, forming a binary system. This allows emission from hot dust near the two protostars to escape without being reprocessed by a cool disc. The result of this is a sustained spike in the emission at  $\lambda\lesssim10\,\mathrm{\upmu m}$. Another peculiar spectrum is observed in Core B (see the last frame of Fig. \ref{fig:core_b_obs}). Here, the system is in a dynamically unstable state and radiation with $\lambda\lesssim10\,\mathrm{\upmu m}$ from dust around one of the objects is more easily observed from an edge on viewing angle than it is face on.

\subsubsection{Viewing angle case study}
\label{sec:viewing_angle}

\begin{figure}
	\includegraphics[width=\columnwidth]{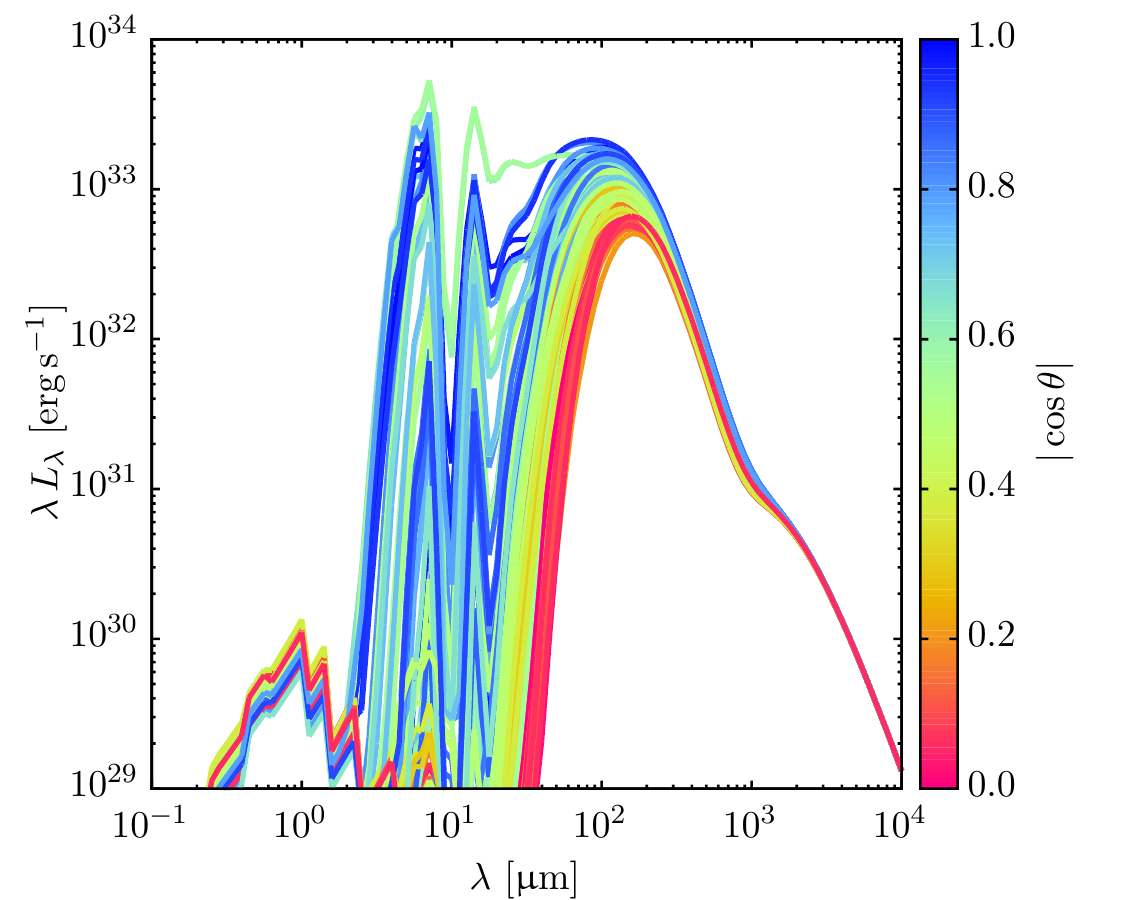}
	\includegraphics[width=\columnwidth]{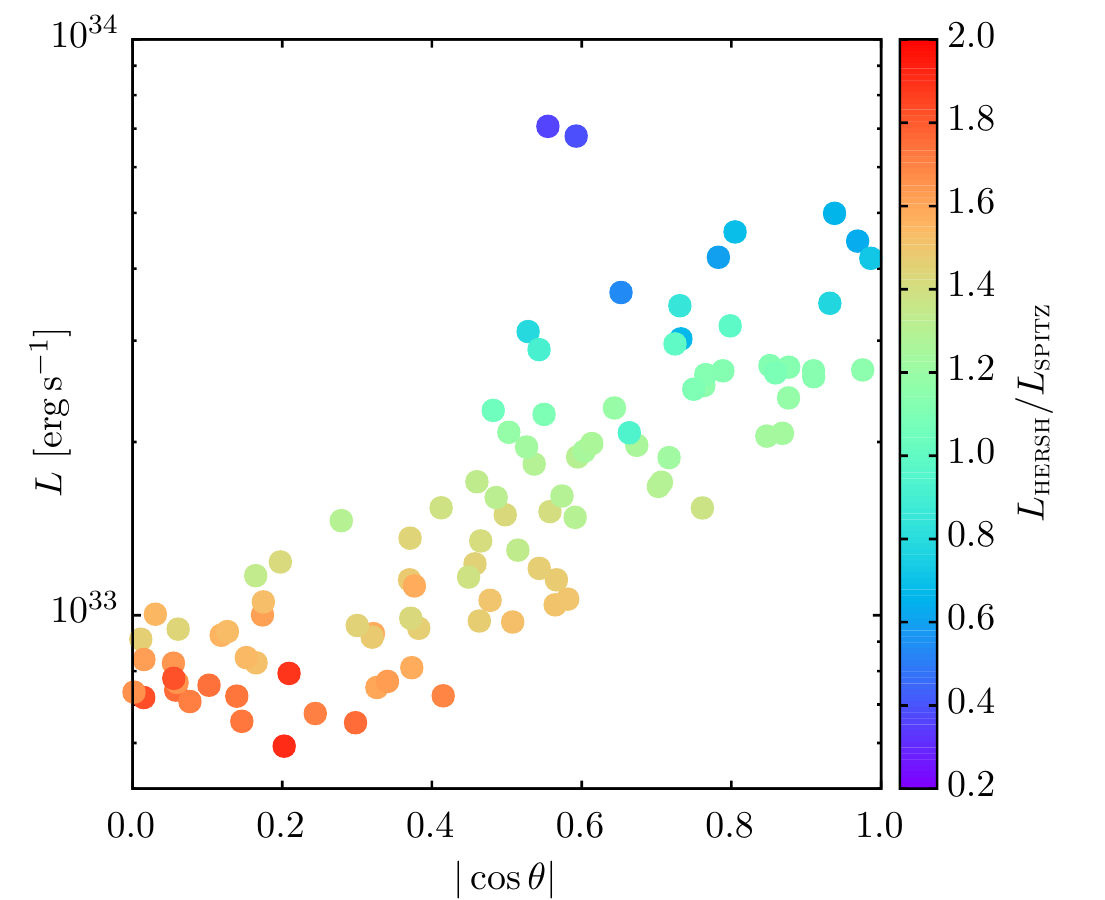}
	\caption{SEDs from Core A at $t=1.7\times10^4\,\mathrm{yrs}$ The top frame shows the full SED with the inclination angle $\theta$ indicated by the colour scale. Note, $|\cos\theta|=1$ corresponds to an approximately face-on view, $\cos\theta=0$ corresponds is an approximately edge-on view. The bottom frame shows the apparent luminosity as a function of inclination angle. Here, the colour scale shows the infrared colour, as defined in Fig. \ref{fig:colour_magnitude}.}
	\label{fig:angle_spectrum}
\end{figure}

\changes{Once a disc has formed, the apparent luminosity of the protostars usually varies strongly with viewing angle. Here, we perform a case study by examining the SEDs of a snapshot (Core A;  $t=1.7\times10^4\,\mathrm{yrs}$), shown in Fig. \ref{fig:angle_spectrum}. A face-on intensity map of this object is shown in Fig. \ref{fig:core_a_obs}. First we identify the principle axes of the system by calculating the inertia tensor for the ensemble of SPH and sink particles. For a flatted rotating object (e.g., a disc), the eigenvector with the largest eigenvalue roughly corresponds to the rotation axis. We define the inclination angle $\theta$ as the angle between the this direction and SED viewing angle.
}

\changes{
We show that, similar to single protostars, the apparent luminosity of the protostellar system is strongly correlated with the viewing angle. In this case, the luminosity when viewed face-on is up to an order of magnitude greater than that when viewed edge-on. There is some scatter in the luminosity-angle relationship which is most likely due to (i) non-axisysmmetric variations in the plane of the system, and (ii) possible deviations between the principle axis and the \emph{true} axis of rotation. We find that the colour of the system is also strongly correlated with the viewing angle. For observed objects where the apparent luminosity, colour and inclination angle are known, this provides a useful tool for estimating its average (or \emph{true}) luminosity.
}

\subsection{Comparison of radiative transfer methods}

\changes{The SW07 radiative transfer approximation and the Monte Carlo approach used by \textsc{spamcart} differ considerably. We discuss these differences in detail in Appendix \ref{sec:rad_trans_comp}. In summary, the SW07 method usually calculates temperatures similar to \textsc{spamcart}. The main exception is within the inner $70\,\mathrm{au}$ of protostellar discs. Here, the SW07 method calculates temperatures which are roughly an order of magnitude greater than those from \textsc{spamcart}. Analytical and numerical calculations of disc fragmentation \citep[e.g.][]{WS06,C09,SW09a} strongly suggest that disc fragmentation should occur outside this radius. We further note that the high inner-disc temperatures can only suppress fragmentation in these regions. The hypothesis that most brown dwarfs and very low mass stars are formed via disc fragmentation \citep[e.g.][]{LWHSW14,LWHSW14b} is therefore \emph{not} undermined by the inaccuracy of the temperature calculation in the SW07 method.
}

\section{Discussion}
\label{sec:discussion}

The SPH simulations here, along with other work \citep[e.g.,][]{LWHSW14,LWHSW14b} challenge the notion that solar type and low mass stars have a simple evolutionary progression from prestellar cores to main sequence stars. Under the conventional model, a prestellar core collapses to form a single class 0 protostar \citep[e.g.,][]{AWB00}. A circumstellar disc forms and material steadily accretes on to the protostar through the class I, class II and class III phases, after which the star progresses onto the main sequence \citep[e.g.,][]{L99}. We suggest that, since cores are turbulent, they frequently fragment into multiple objects. Once this occurs, $N$-body processes and anisotropic accretion can produce protostellar multiple systems with separations, $s\gtrsim1\,\mathrm{au}$, and highly varied morphologies. This makes interpreting observations difficult as there is a significant chance that one or more protostars will be in the same telescope beam. This is even true for powerful interferometers such as the Very Large Array (VLA) and the Atacama Large Millimetre Array (ALMA).

\subsection{SED fitting methodology}

It may be possible to use simulations to address difficulties in fitting model protostars to observations. For simplicity, we will only discuss protostellar SEDs. If we consider a protostellar system with parameters $\boldsymbol{\theta}$ (e.g., number of objects, primary/secondary luminosity, primary/secondary disc mass, circumstellar disc mass, etc.), the posterior probability of $\boldsymbol{\theta}$, given some observed data $D$, is given by Bayes' theorem,
\begin{equation}
	P(\boldsymbol{\theta}|D)=\frac{P(D|\boldsymbol{\theta})\,P(\boldsymbol{\theta})}{P(D)}\,.
\end{equation}
Here, $P(D|\boldsymbol{\theta})$ is the likelihood of $D$, given $\boldsymbol{\theta}$, $P(\boldsymbol{\theta})$ is the \emph{a priori} probability of $\boldsymbol{\theta}$ and $P(D)$ is a normalisation constant such that integral of $P(\boldsymbol{\theta}|D)$ over all $\boldsymbol{\theta}$ is equal to one. Two difficulties arise from this analysis. First, the number of dimensions in $\boldsymbol{\theta}$-space can be arbitrarily large. For example, \citet{R17} defines a single protostar using a parameter space with at least 7 dimensions. This scales superlinearly with number of possible objects you permit in a multiple system, e.g., a quadruple system under this model has at least 28 parameters. If the prior distribution is uninformative (usually uniform or log-uniform between upper and lower limits along each dimension), then the number of sampling points required to cover the parameter space increases geometrically with each additional object. Second, fits to protostellar SEDs are often degenerate. A notable example of this is the degeneracy between disc mass and temperature caused by the Rayleigh-Jeans tail of the Planck function.

Both of these issues can be addressed by performing a large number of core simulations, with initial conditions representative of star forming regions \citep[e.g.,][]{LWHSW14}, and using the results as an informative prior. First, we run $m$ simulations and take $n$ snapshots from each. We post-process each snapshot $p$ times from different random viewing directions. This produces $N=m\,n\,p$ synthetic observations, each with $\boldsymbol{\theta}_j$ which is measured from the snapshot. Here, $\boldsymbol{\theta}_j$ is effectively sampled from a prior probability distribution defined by the suite of simulations. \changes{A posterior probability distribution can be estimated by calculating $\chi^2$ likelihoods for each $\boldsymbol{\theta}_j$, which in turn can be used to estimate parameter expectation values.}

Performing an analysis using this method eliminates the need to explicitly set up an arbitrarily large parameter space with an associated prior distribution. Furthermore, the simulation parameters are hyperparameters (i.e. parameters of the prior distribution) of the posterior distribution. This helps to lift degeneracies in SED fits; when the likelihood is unable to discriminate between degenerate combinations of parameters, the posterior is determined by the simulations instead. \changes{However the hyperparameters are likely to have an affect on the posterior. The simulations presented here have caveats and limitations (discussed in Section \ref{sec:caveats}) which would inevitably affect this analysis. We therefore stress that this methodology must be repeated as the state of the art (e.g., sophistication of numerical models and available computing resources) progresses. Furthermore, statements from any Bayesian analysis are entirely dependent on the model used. We envisage that comparing the results of this analysis with multiple models would have one of two main outcomes. First, ideally, the results may vary little when different models are used. This allows us to confidently place constraints on the physical properties of protostars. Second, the results may deviate when different models are used. While this is less ideal than the first case, it still provides a means of quantitatively comparing the outcomes of different models, given observations. This may aid further research in eliminating or reconciling various models.}

Such an analysis requires a much larger suite of simulations than that presented here. This is numerically feasible, given that each simulation \emph{unit}, i.e., an SPH simulation plus $4\times10^4$ synthetic observations, requires $\sim4\times10^4$ CPU hours on current hardware. Scaling this up to, say, 100 units is easily achievable over a year with large computing clusters, e.g., the Distributed Research utilising Advanced Computing (DiRAC) supercomputing facilities.

\subsection{Caveats}
\label{sec:caveats}

We have presented a sophisticated suite of numerical simulations which follow the observational properties of protostellar multiple systems over time. However, there are some important caveats we must address.


\changes{The resolution of the SPH simulations matches the requirements of modelling the hydrodynamics and self gravity of star formation. These do not necessarily match the requirements of modelling full radiative transfer through the ISM. The method presented here conserves photons \emph{exactly} during the temperature calculations. However, the ray-tracing method used to generate the synthetic observations is approximate. In regions with extreme temperature gradients (e.g., in the vicinity of sinks), lack of resolution can cause the ray-trace to overestimate the intensity (and hence luminosity) by 30--50\%.}

\changes{The interstellar radiation field is assumed to be isotropic and unattenuated by the core's parent molecular cloud. It is therefore only a rough estimate of the external heating. In nature, the strength of this term varies from region to region. For deeply embedded starless cores, the radiation field is attenuated by dust. Conversely, starless cores near high mass stars have a greater degree of heating. However, once a protostar forms, it quickly becomes the dominant heating source for the core. Here, effect of the background on integral quantities, such as luminosity and colour, is negligible.}

We have not modelled jets in the SPH simulations. This means that there are no cavities in the core envelopes above and below the protostellar poles. As a result, there may be less mid-infrared energy in the face-on system SEDs than there should be. A sub-grid model of protostellar jets (P. Rohde and S. Walch, private communication) is in development for the \textsc{gandalf} SPH code \citep{HR13} and will be included in future studies.

We use a single dust model throughout the density field which does not account for polycyclic aromatic hydrocarbons (PAHs) or large coagulated dust grains with icy mantles \citep[e.g.,][]{OH94}. We also assume that the dust-to-gas mass ratio is fixed outside dust sublimation regions. Modelling the evolution and dynamics of dust is complicated \citep[e.g.,][]{RDL17,HR13} and will need to be considered in future.

The SWH11 model assumes that a protostellar photosphere is a blackbody with a variable luminosity and a fixed radius. The luminosity is sensitive to the sub-grid accretion model, the parameters of which have weak observational constraints (see SWH11 and references therein). When applying these models to SED fitting, the luminosity should not be interpreted as a proxy for stellar mass or accretion rate.

\section{Summary}
\label{sec:summary}

We have presented synthetic observations of three protostellar multiple systems at different points in time, viewed from different angles. The gas dynamics of the systems are modelled using SPH and the synthetic observations are generated using the \textsc{spamcart} post-processing Monte Carlo radiative transfer code. These are the first synthetic observations of protostars which take into account episodic accretion and multiplicity. We note that we have modified \textsc{spamcart} so that it now accounts for dust scattering and extremely optically thick objects, such as protostellar discs. The \textsc{spamcart} code makes full use of the detailed SPH particle distribution and we are able to resolve spatial scales $r\lesssim0.1\,\mathrm{au}$, which is equivalent to the dust sublimation radii of protostars.

There is a growing amount of observational and numerical evidence indicating that many stars begin their lives in multiple systems. This suggests that studies which attempt to infer the physical properties of observed protostars using single star models may be inappropriate. We propose a Bayesian methodology, based on an expansion of the calculations presented here, that can be used to infer the physical properties of potentially multiple protostellar systems. We plan to undertake this project in the near future.

\bibliographystyle{mn2e}
\bibliography{refs}

\section*{Acknowledgements}

We thank the anonymous referee for their constructive comments which improved the clarity of this paper. OL and APW gratefully acknowledge the support of a consolidated grant (ST/K00926/1) from the UK STFC. Calculations were performed using the computational facilities of the Advanced Research Computing @ Cardiff (ARCCA) Division, Cardiff University.

\appendix

\section{Numerical treatment of dust radiative transfer}

\subsection{Dust properties}
\label{sec:dust_properties}

We use the values of $\kappa_\lambda$, $\sigma_\lambda$ and mean scattering cosine $g_\lambda$ derived by \citep{LD01}. The dust is a mixture of carbonaceous and amorphous silicate grains, with a size distribution following \citet{WD01} at $R_v=5.5$, and a dust-to-gas mass ratio of one percent. We simplify the treatment of dust scattering by approximating the scattering phase function $p_\lambda(\cos\theta)$ with a linear combination of isotropic scattering and pure forward or backward scattering. Here,
\begin{equation}
	p_\lambda(\cos\theta)=
	\begin{cases}
		\frac{1}{2}\,(1-g_\lambda)+2\,g_\lambda\,\delta(\cos\theta-1), & g_\lambda>0\,;\\
		\frac{1}{2}\,(1+g_\lambda)-2\,g_\lambda\,\delta(\cos\theta+1), & g_\lambda<0\,;\\
		\frac{1}{2}, & g_\lambda=0\,,
	\end{cases}
	\label{eqn:phase_function}
\end{equation}
where $-1\leq\cos\theta\leq1$. This phase function can be implemented in a numerical code when we note that the delta function component represents a modification to the local scattering mean free path $l_\lambda$, i.e.,
\begin{equation}
	l'_\lambda=\sum\limits_{n=0}^\infty l_\lambda\,g_\lambda^n=\frac{l_\lambda}{1-g_\lambda}\,.
\end{equation}
We may therefore model scattering as a completely isotropic process by defining a replacement set of optical properties:
\begin{equation}
	\begin{split}
		\sigma'_\lambda&\equiv(1-g_\lambda)\sigma_\lambda\,;\\
		\kappa'_\lambda&\equiv\kappa_\lambda\,;\\
		g'_\lambda&\equiv0\,.
	\end{split}
\end{equation}

\subsection{Dust sublimation}
\label{sec:dust_sub}

We assume that dust sublimation occurs at temperatures above $T_\textsc{sub}=1000\,\mathrm{K}$. We reduce the dust-to-gas mass ratio for each particle by a factor $a_i$, which is calculated from $N_\textsc{neib}$ neighbours,
\begin{equation}
	a_i=\sum_{j=1}^{N_\textsc{neib}}\frac{m_j}{h_i^3\,\rho_i}\,w(\boldsymbol{r}_{ij})\,\max[H(T_\textsc{sub}-T_j),\epsilon]\,.
\end{equation}
Here, $h$ is the particle smoothing length, $\rho$ is the particle density, $w(\boldsymbol{r})$ is the smoothing kernel function, $H(T)$ is the Heaviside step function and $\epsilon$ is the smallest floating point number where $1+\epsilon>1$ ($\epsilon\simeq2\times10^{-16}$ for 64 bit floating point variables). This ensures that $a_i\cong1$ if $T_j<T_\textsc{sub}$ is true for all neighbouring particles and $a_i\cong\epsilon$ if $T_j\ge T_\textsc{sub}$ for all neighbours. 

\subsection{Modified random walk}
\label{sec:mrw}

In very optically thick regions, such as the mid-planes of discs, the mean free path of the luminosity packets may be several orders of magnitude smaller than the local particle smoothing length. In these situations, following the full trajectory of the packet is prohibitively expensive. We address this by using the modified random walk (MRW) algorithm, originally presented by \citet{MDD09} and simplified by \citet{R10}. 

In its original form, the MRW moves packets through grid cells which have uniform Planck inverse mean volume opacity, $\bar{\chi}_\textsc{vol}=\rho\,\bar{\chi}_\textsc{inv}(T)$, where
\begin{equation}
	\bar{\chi}_\textsc{inv}(T)=\frac{\int\limits_0^\infty B_\lambda(T)\,\mathrm{d}\lambda}{\int\limits_0^\infty \chi_\lambda^{-1}B_\lambda(T)\,\mathrm{d}\lambda}\,.
\end{equation}
Here, a packet which has diffused to some distance $r_\textsc{mrw}$ from its origin $\boldsymbol{r_0}$, has a total random walk length
\begin{equation}
	s_\textsc{rw}=-\ln y\,\left(\frac{r_\textsc{mrw}}{\uppi}\right)^2\,3\,\bar{\chi}_\textsc{vol}\,,
	\label{eqn:mrw_step}
\end{equation}
where $y$ is found by numerically inverting the equation,
\begin{equation}
	\mathcal{R}=2\,\sum\limits_{n=1}^\infty (-1)^{n+1}\,y^{n^2}\,,
\end{equation}
and $\mathcal{R}$ is randomly drawn from the uniform distribution in the interval $[0,1]$. In practice, a packet is moved a distance $r_\textsc{mrw}$ (within the same cell) from $\boldsymbol{r_0}$ to $\boldsymbol{r_1}$ with a random isotropic direction $\hat{\boldsymbol{r}}$. The distance $s_\textsc{rw}$ is used instead of $r_\textsc{mrw}$ to update the energy absorption rate. In order to apply this method here, we must (i) formulate a column density equivalent of Eqn. \ref{eqn:mrw_step} and (ii) and account for inhomogeneity of $\bar{\chi}_\textsc{vol}(\boldsymbol{r})$ within a smoothed particle ensemble.

During a MRW step, a packet moves some number of mean free paths $\tau_\textsc{mrw}$. The relationship between $\tau_\textsc{mrw}$ and distance $r_\textsc{mrw}$ is defined:
\begin{equation}
	\tau_\textsc{mrw}=\int\limits_C\bar{\chi}_\textsc{vol}(\boldsymbol{r})\,\mathrm{d}s\,,
	\label{eqn:mrw_tau}
\end{equation}
where $C$ is the line segment,
\begin{equation}
	\boldsymbol{r}=\boldsymbol{r_0}+s\,\hat{\boldsymbol{r}}\,,\quad0\leq s\leq r_\textsc{mrw}\,.
\end{equation}
Here, a solution for $r_\textsc{mrw}$ given $\tau_\textsc{mrw}$ is found using the modified Newton-Raphson method detailed by LW16.

We perform a MRW step from $\boldsymbol{r_0}$ to $\boldsymbol{r_1}$ if the local mean free path $\bar{\chi}_\textsc{vol}^{-1}(\boldsymbol{r_0})$ is a factor of a few less than the local smoothing length $h(\boldsymbol{r_0})$. Our choice of $\tau_\textsc{mrw}$ depends on the local environment:
\begin{equation}
	\tau_\textsc{mrw}=\min\left(h(\boldsymbol{r_0})\,\bar{\chi}_\textsc{vol}(\boldsymbol{r_0}),\frac{\bar{\chi}_\textsc{vol}(\boldsymbol{r_0})^2}{2\,\lVert\nabla\bar{\chi}_\textsc{vol}(\boldsymbol{r_0})\rVert}\right)\,.
\end{equation}
The first term on the right hand side ensures that $r_\textsc{mrw}\lesssim h(\boldsymbol{r_0})$; the second terms attempts to keep $r_\textsc{mrw}$ finite for all values of $\hat{\boldsymbol{r}}$. The quantities $h(\boldsymbol{r_0})$, $\bar{\chi}_{\textsc{vol}}(\boldsymbol{r}_0)$ and $\nabla\bar{\chi}_\textsc{vol}(\boldsymbol{r}_0)$ are estimated using SPH scatter calculations. We pick a random $\hat{\boldsymbol{r}}$, calculate $r_\textsc{mrw}$ and move the packet to position $\boldsymbol{r_1}=\boldsymbol{r_0}+r_\textsc{mrw}\,\hat{\boldsymbol{r}}$. The energy absorption and scattering rates of particles intersected by the path (see Eqn. \ref{eqn:dust_abs} and \ref{eqn:dust_sca}) are updated,
\begin{equation}
	\begin{split}
		\dot{A}_i&\to\dot{A}_i+\mathcal{W}_i\left(\frac{s_\textsc{rw}}{r_\textsc{mrw}}\right)\frac{\ell_j}{m_i}\kappa_\textsc{p}(T_i)\,\varsigma_{ij}\,,\\
		\dot{S}_{i\lambda}\,\mathrm{d}\lambda&\to\dot{S}_{i\lambda}\,\mathrm{d}\lambda+\mathcal{W}_i\left(\frac{s_\textsc{rw}}{r_\textsc{mrw}}\right)\frac{\ell_j}{m_i}\sigma_\textsc{p}(T_i,\lambda,\mathrm{d}\lambda)\,\varsigma_{ij}\,.
	\end{split}
\end{equation}
Here, $\sigma_\textsc{p}(T,\lambda,\mathrm{d}\lambda)$ is the partial Planck mean mass scattering coefficient,
\begin{equation}
	\sigma_\textsc{p}(T,\lambda,\mathrm{d}\lambda)=\frac{\int\limits_\lambda^{\lambda+\mathrm{d}\lambda} \sigma_{\lambda'}\,B_{\lambda'}(T)\,\mathrm{d}\lambda'}{\int\limits_0^\infty B_{\lambda'}(T)\,\mathrm{d}\lambda'}\,.
\end{equation}
The expression $(s_\textsc{rw}/r_\textsc{mrw})$ is found by rearranging Eqn. \ref{eqn:mrw_step} and replacing $r_\textsc{mrw}\,\bar{\chi}_\textsc{vol}$ with $\tau_\textsc{mrw}$ from Eqn \ref{eqn:mrw_tau}, i.e.,
\begin{equation}
	\left(\frac{s_\textsc{rw}}{r_\textsc{mrw}}\right)=-\ln y\,\frac{3\,\tau_\textsc{mrw}}{\uppi^2}\,.
\end{equation}
We include a weighting term $\mathcal{W}_i$ which takes into account the variation in volume opacity along the random walk. Here,
\begin{equation}
	\mathcal{W}_i=\rho_i\,\bar{\chi}_\textsc{inv}(T_i)\,\frac{\sum\limits_k\bar{\chi}_\textsc{p}(T_k)\,\varsigma_{kj}}{\sum\limits_k\rho_k\,\bar{\chi}_\textsc{inv}(T_k)\,\bar{\chi}_\textsc{p}(T_k)\,\varsigma_{kj}}\,,
\end{equation}
where $k$ are the indices of all intersected particles and $\chi_\textsc{p}(T)$ is the Planck mean mass opacity coefficient. This term makes sure that the share of $s_\textsc{rw}$ assigned to each particle is proportional to the volume opacity of each particle. In regions with roughly homogeneous volume opacity, $\mathcal{W}_i$ is approximately equal to one.

\section{Kirchoff's law benchmark}
\label{sec:kirchoff}

\begin{figure*}
	\centering
	\includegraphics[height=0.27\textwidth]{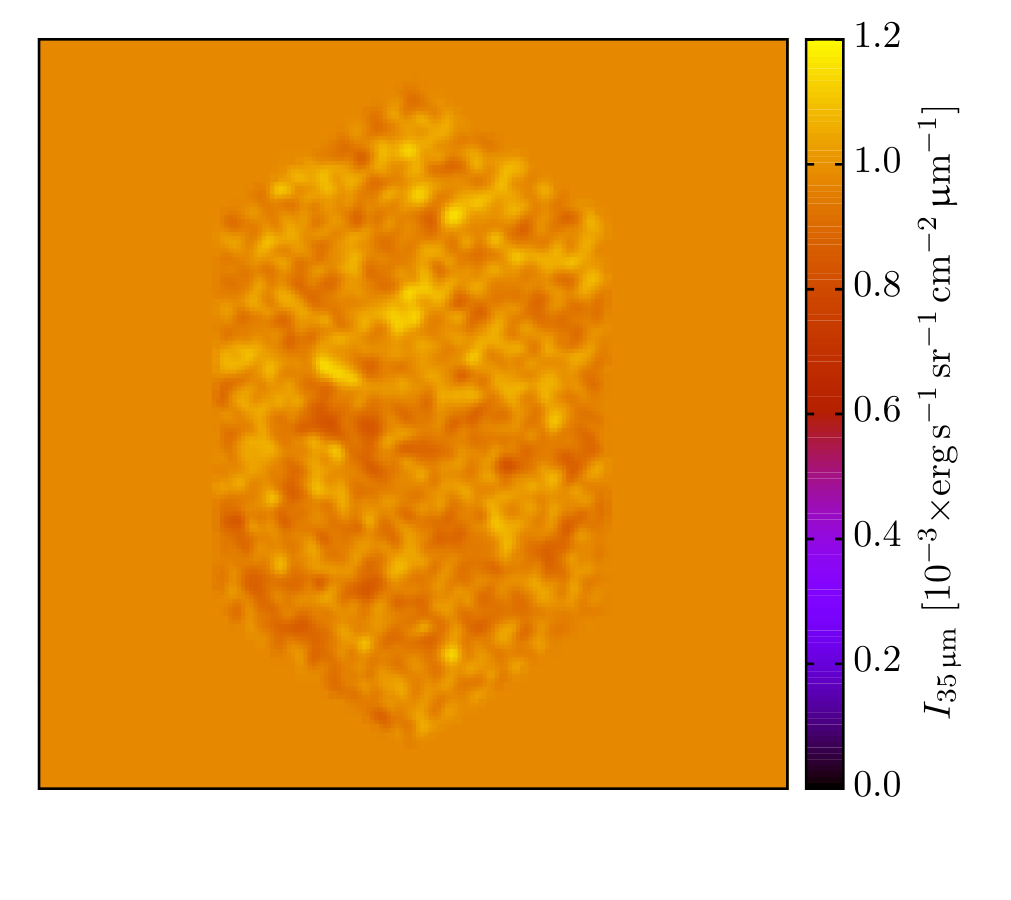}
	\includegraphics[height=0.27\textwidth]{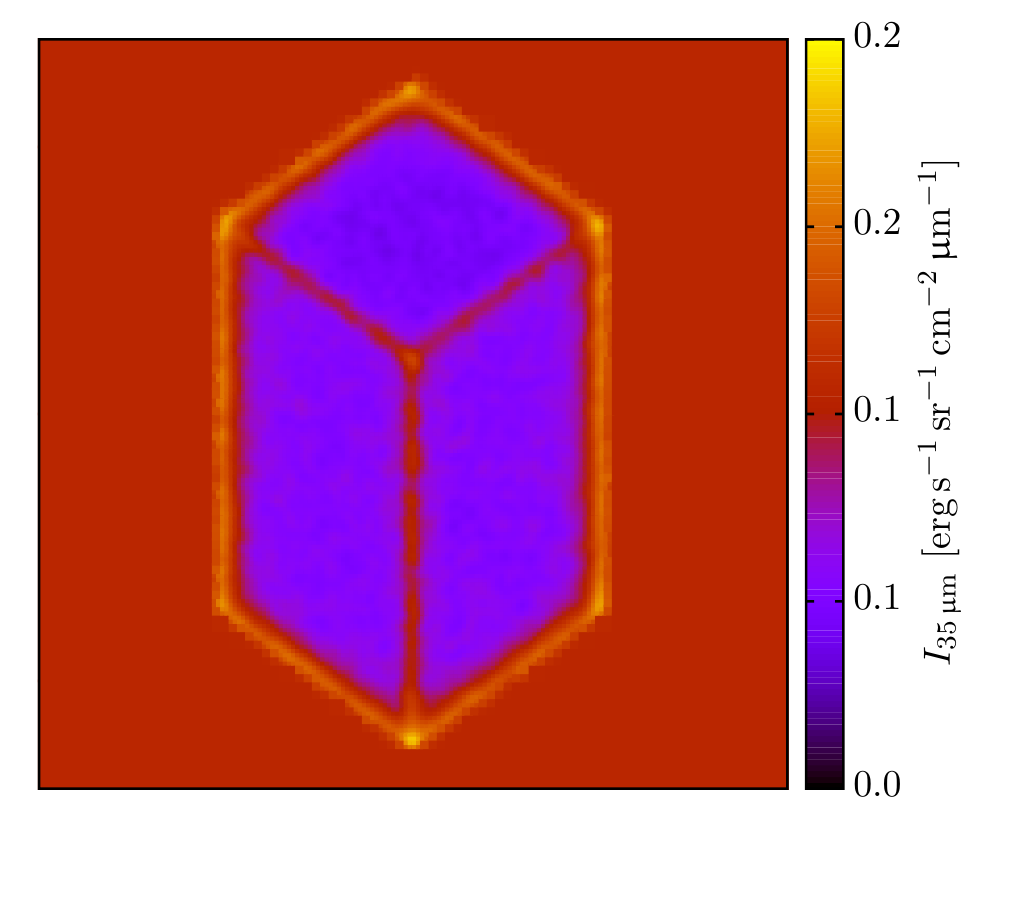}
	\includegraphics[height=0.27\textwidth]{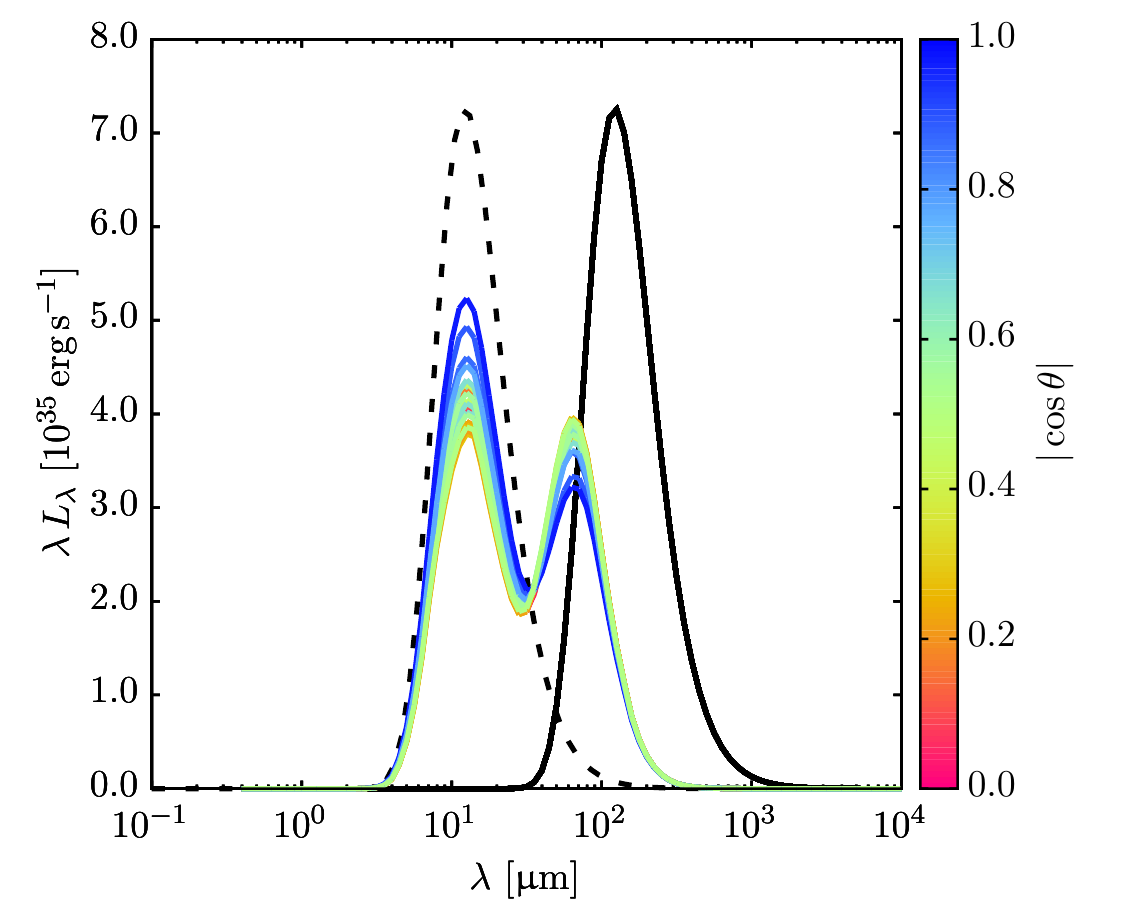}
	\caption{The left and centre frames show the $35\,\upmu\mathrm{m}$ intensity maps of the cuboid in Field A (left) and Field B (centre), offset from its principle axes by 45$^\circ$. The edge-length of both frames is $1300\,\mathrm{au}$. The right frame shows the SED of the cuboid in Field A (black solid lines) and Field B (multicoloured lines) along with a $300\,\mathrm{K}$ blackbody SED diluted by a factor of $10^{-4}$. The colour scale shows the angle between the long axis of the cuboid and the direction normal to the viewing plane. When the cuboid is viewed along the long axis, $|\cos\theta|=1$; when the long axis is parallel to the plane of the sky, $\cos\theta=0$. Note that the solid black line is actually 30 overlaid SEDs which are indistinguishable on this plot.} 
	\label{fig:kirchoff}
\end{figure*}

An object in radiative equilibrium with a uniform blackbody background radiation field, $I_{\lambda\textsc{bg}}=B_\lambda(T_\textsc{bg})$, should have temperature $T_\textsc{bg}$. This becomes apparent by stating that the energy absorption rate at on the surface of the object $\mathcal{A}$ equal to the emission rate $\mathcal{E}$, where
\begin{equation}
	\begin{split}
		\mathcal{A}&=\uppi\int\limits_A\int\limits_{\lambda=0}^\infty I_{\lambda\textsc{bg}}\,\alpha_{\lambda}\,\mathrm{d}\lambda\,\mathrm{d}A\,;\\
		\mathcal{E}&=\uppi\int\limits_A\int\limits_{\lambda=0}^\infty B_\lambda(T_\textsc{ob})\,\epsilon_{\lambda}\,\mathrm{d}\lambda\,\mathrm{d}A\,.
	\end{split}
\end{equation}
Here, $\alpha_{\lambda}$ and $\epsilon_{\lambda}$ are the absorptivity and emissivity respectively, and $T_\textsc{ob}$ is the surface temperature of the object. Kirchoff's law of thermal radiation states $\alpha_{\lambda}=\epsilon_{\lambda}$. If $I_{\lambda\textsc{bg}}=B_\lambda(T_\textsc{bg})$, then $T_\textsc{ob}$ must equal $T_\textsc{bg}$ in order to satisfy the condition $\mathcal{A}=\mathcal{E}$. Furthermore, the intensity at the object's surface $I_{\lambda\textsc{ob}}$ is equal to $I_{\lambda\textsc{bg}}$ and the object is indistinguishable from the background. Note that in this example we have omitted the directional dependence from the terms to simplify the equations; the above still holds if directional dependence is included. If we dilute the background radiation field, i.e., $I_{\lambda\textsc{bg}}=d\,B_\lambda(T_\textsc{bg})$, where $d<1$, then $T_\textsc{ob}<T_\textsc{bg}$. This will result in short-wavelength radiation being absorbed and reemitted at longer wavelengths. The effect of this on the SED of an object is difficult to predict quantitatively without numerical calculations, however we should see a reduction in the background SED and a second peak at a longer wavelengths than the background.

We re-examine the test of Kirchoff's law presented by LW16. The original test demonstrated that \textsc{spamcart} adheres to Kirchoff's law when a spherically-symmetric distribution of particles is irradiated. Here, we perform a similar test, but with a non-spherically symmetric distribution of particles. We construct a cuboid of $2\times10^5$ SPH particles, with total mass $2\,\mathrm{M_\odot}$ and dimensions $2000\,\mathrm{au}\times\,1000\,\mathrm{au}\times\,1000\,\mathrm{au}$. This is illuminated by two background fields: Field A, which has the SED of an undiluted $30\,\mathrm{K}$ blackbody; and Field B with has the SED of a $300\,\mathrm{K}$ blackbody, diluted by a factor of $10^{-4}$. In each case, the field is simulated using $10^6$ luminosity packets and five iterations. The SEDs and intensity maps are constructed for 30 random viewing angles.

Fig. \ref{fig:kirchoff} shows example intensity maps and the SEDs of the illuminated cuboids. The two maps show the intensity at $35\,\mathrm{\upmu m}$. With Field A, the surface of the cuboid has the same intensity as the background, modulo fluctuations of about 20\%. At these wavelengths, the path through the cuboid corresponds to 30 to 60 optical depths. Most of the emission associated with the cuboid is generated from the outermost layer of SPH particles only. At wavelengths closer to the peak emission -- say, $200\,\mathrm{\upmu m}$ -- the intensity fluctuations relative to the blackbody background are $\ll1\%$. The SEDs for all 30 viewing angles are indistinguishable from a $30\,\mathrm{K}$ blackbody. With Field B, the cuboid casts a silhouette over the background at $35\,\mathrm{\upmu m}$. The edges of the cuboid are warmer than the faces (the corners warmer still) because they are heated from multiple directions. The SEDs show that the cuboid reduces the flux arriving from the background radiation field. This is most pronounced when the long axis of the cuboid is parallel to the image plane. The absorbed emission is reemitted by the cuboid at $\sim70\,\mathrm{\upmu m}$, where it is noticeably brighter than the background. In summary, the numerical results from \textsc{spamcart} are in agreement with our predictions based on Kirchoff's law.

\section{Comparison of radiative transfer methods}
\label{sec:rad_trans_comp}

\begin{figure}
	\centering
	\includegraphics[width=\columnwidth]{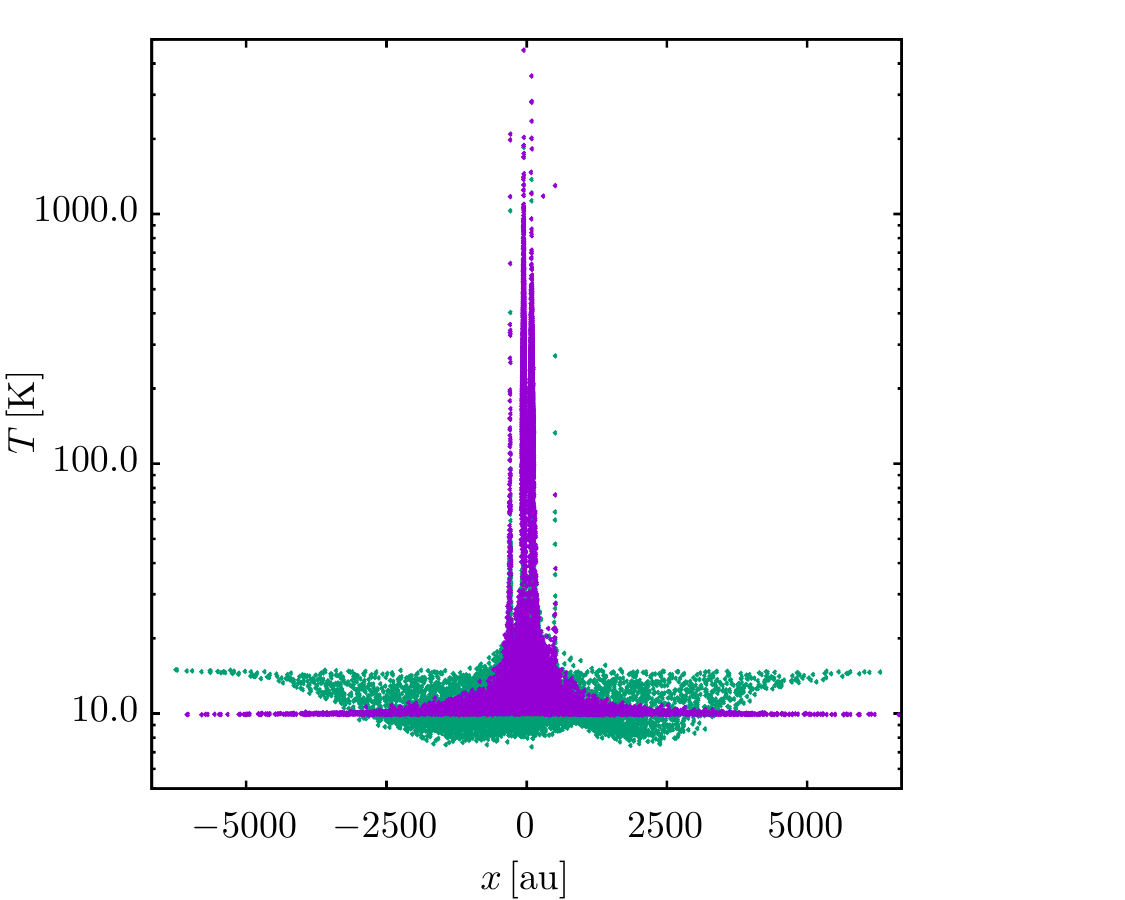}
	\includegraphics[width=\columnwidth]{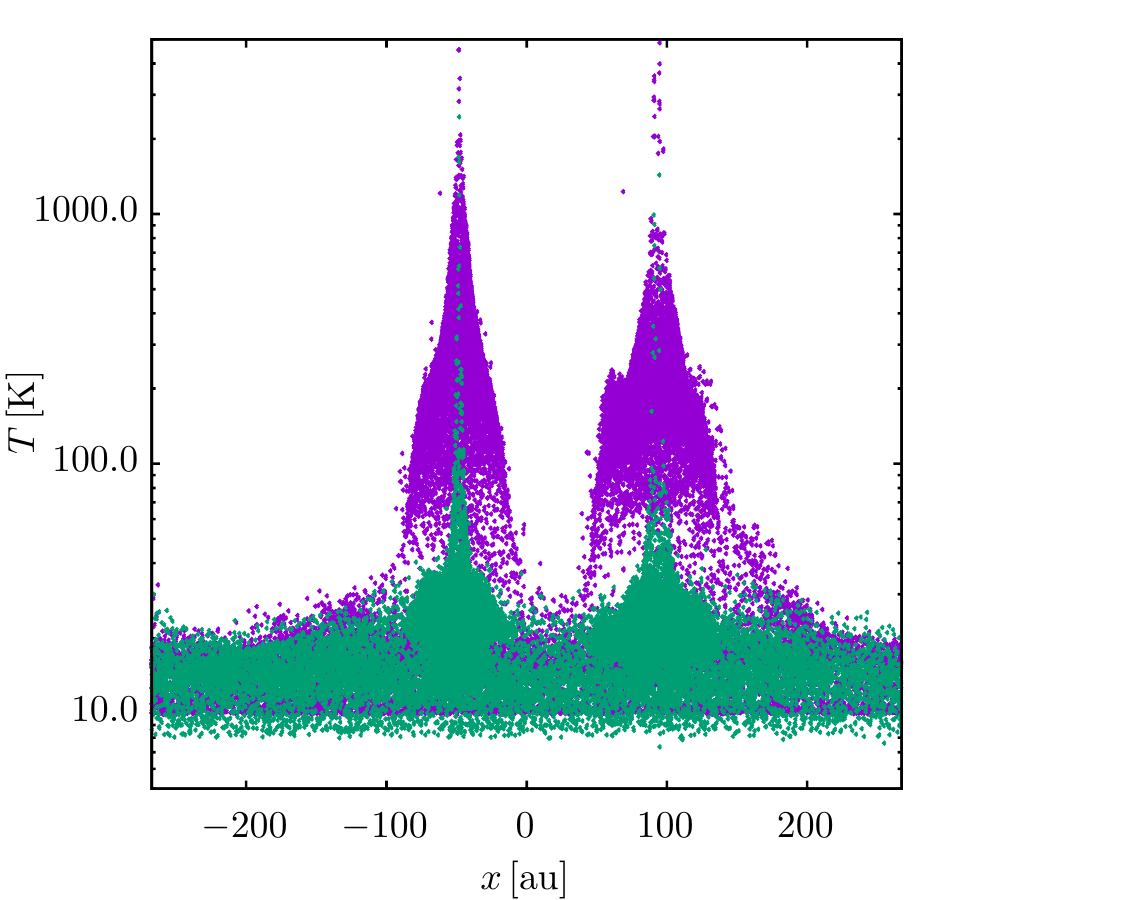}
	\caption{Temperature vs $x$-coordinate for Core A at $t=1.7\times10^4\,\mathrm{yrs}$. The purple points show the temperatures calculated during the SPH simulations using the SW07 method. The green points show the temperatures calculated using \textsc{spamcart}. The two frames show the distribution at two different spatial scales.}
	\label{fig:rad_trans_comp}
\end{figure}

The Monte Carlo radiative transfer calculations presented here differ significantly from the SW07 radiative transfer approximation used in the SPH simulations. Here, we provide a brief discussion of these differences. The SW07 method approximates radiative cooling and heading rates by assuming that each particle is embedded in a polytropic pseudo-cloud. The physical properties of the pseudo-cloud are tuned to the density, temperature and gravitational potential at the particle position. This is used to calculate a mass-weighted exit optical depth from the pseudo-cloud, which is in turn used to estimate the amount of radiative shielding affecting the heating and cooling rate of each particle. Table \ref{tab:SW07} gives a brief comparison of the physical assumptions used in \textsc{spamcart} and the SW07 method.

\begin{table}
	\centering
	\begin{tabular}{|p{0.2\columnwidth}|p{0.33\columnwidth}|p{0.33\columnwidth}|}
		\hline
		& \textsc{spamcart} & SW07 \\
		\hline
		\scriptsize{Time dependence.} & \scriptsize{Equilibrium calculation.} & \scriptsize{Assumes thermalisation timescale.}\\
		\hline
		\scriptsize{Geometric assumptions.} & \scriptsize{None.} & \scriptsize{Spherical symmetry.}\\
		\hline
		\scriptsize{Heating sources.} & \scriptsize{Radiation from stars and background radiation field, scattered and absorbed/reemitted by dust.} & \scriptsize{Background temperature. Quartic sum of interstellar backgound ($T=10\,\mathrm{K}$) and local stellar temperature ($T\propto r^{-2}$). Heating from $P\,\mathrm{d}V$ work done.}\\
		\hline
		\scriptsize{Opacity sources.} & \scriptsize{Dust \citep{LD01,WD01}.} & \scriptsize{Dust and gas \citep{BL94}.}\\
		\hline
	\end{tabular}
	\caption{A brief summary of the physical assumptions used in the \textsc{spamcart} code compared with those used in the SW07 method.}
	\label{tab:SW07}
\end{table}

Fig. \ref{fig:rad_trans_comp} shows the distribution of particle temperatures from a snapshot of Core A. Here, we have plotted temperatures calculated using \textsc{spamcart} and using the SW07 method. We note a number of differences in the temperatures. First, for the material within $70\,\mathrm{au}$ of sinks (i.e., accretion discs), the SW07 temperatures can be roughy an order of magnitude greater than the \textsc{spamcart} temperatures. This is because the SW07 method assumes spherical symmetry and therefore material can not cool as efficiently as it physically should in the presence of strong density gradients. Second, the distribution of temperatures in the inner core envelope ($R\lesssim3000\,\mathrm{au}$) is more varied in the \textsc{spamcart} calculations than SW07. This is due to \textsc{spamcart} fully modelling anisotropies in the local radiation field (plus a small level of statistical noise). Finally, The \textsc{spamcart} temperature at the core boundary is a roughly $15\,\mathrm{K}$ whereas the SW07 temperature is $10\,\mathrm{K}$. This is due to differences in the \citet{PS08} radiation field used in \textsc{spamcart} and the assumed $10\,\mathrm{K}$ radiation field used by SW07.

\label{lastpage}

\end{document}